\documentclass{pasj00}

\begin{document}
\SetRunningHead{Hirota et al.}{VERA astrometry of IRAS~22198+6336 in L1204G}
\Received{2008/04/28}
\Accepted{2008/08/04}

\title{Astrometry of H$_{2}$O Masers in Nearby Star-Forming Regions with VERA. 
III. IRAS~22198+6336 in L1204G}

\author{
Tomoya \textsc{Hirota},\altaffilmark{1,2}
Kazuma \textsc{Ando},\altaffilmark{3} 
Takeshi \textsc{Bushimata},\altaffilmark{1,4}
Yoon Kyung \textsc{Choi},\altaffilmark{1} \\
Mareki \textsc{Honma},\altaffilmark{1,2} 
Hiroshi \textsc{Imai},\altaffilmark{5} 
Kenzaburo \textsc{Iwadate},\altaffilmark{6}
Takaaki \textsc{Jike},\altaffilmark{6} \\
Seiji \textsc{Kameno},\altaffilmark{5} 
Osamu \textsc{Kameya},\altaffilmark{2,6} 
Ryuichi \textsc{Kamohara},\altaffilmark{1} 
Yukitoshi \textsc{Kan-ya},\altaffilmark{7} \\
Noriyuki \textsc{Kawaguchi},\altaffilmark{1,2,4} 
Masachika \textsc{Kijima},\altaffilmark{2} 
Mi Kyoung \textsc{Kim},\altaffilmark{1,8} 
Hideyuki \textsc{Kobayashi},\altaffilmark{1,4,6,8} \\
Seisuke \textsc{Kuji},\altaffilmark{6} 
Tomoharu \textsc{Kurayama},\altaffilmark{9}
Seiji \textsc{Manabe},\altaffilmark{2,6} 
Makoto \textsc{Matsui},\altaffilmark{3} \\
Naoko \textsc{Matsumoto},\altaffilmark{2} 
Takeshi \textsc{Miyaji},\altaffilmark{1,4}
Atsushi \textsc{Miyazaki},\altaffilmark{6}
Takumi \textsc{Nagayama},\altaffilmark{3} \\
Akiharu \textsc{Nakagawa},\altaffilmark{5}   
Daichi \textsc{Namikawa},\altaffilmark{3} 
Daisuke \textsc{Nyu},\altaffilmark{3} 
Chung Sik \textsc{Oh},\altaffilmark{1,8} \\
Toshihiro \textsc{Omodaka},\altaffilmark{5} 
Tomoaki \textsc{Oyama},\altaffilmark{1} 
Satoshi \textsc{Sakai},\altaffilmark{6} 
Tetsuo \textsc{Sasao},\altaffilmark{9,10} \\
Katsuhisa \textsc{Sato},\altaffilmark{6} 
Mayumi \textsc{Sato},\altaffilmark{1,8} 
Katsunori M. \textsc{Shibata},\altaffilmark{1,2,4} 
Yoshiaki \textsc{Tamura},\altaffilmark{2,6} \\
Kosuke \textsc{Ueda},\altaffilmark{3} 
and Kazuyoshi \textsc{Yamashita}\altaffilmark{2}
}
\altaffiltext{1}{Mizusawa VERA Observatory, National Astronomical Observatory of Japan, \\
  2-21-1 Osawa, Mitaka, Tokyo 181-8588}
\altaffiltext{2}{Department of Astronomical Sciences, Graduate University for Advanced Studies, \\
  2-21-1 Osawa, Mitaka, Tokyo 181-8588}
\altaffiltext{3}{Graduate School of Science and Engineering, Kagoshima University, \\
  1-21-35 Korimoto, Kagoshima, Kagoshima 890-0065}
\altaffiltext{4}{Space VLBI Project, National Astronomical Observatory of Japan, \\
  2-21-1 Osawa, Mitaka, Tokyo 181-8588}
\altaffiltext{5}{Faculty of Science, Kagoshima University, \\
  1-21-35 Korimoto, Kagoshima, Kagoshima 890-0065}
\altaffiltext{6}{Mizusawa VERA Observatory, National Astronomical Observatory of Japan, \\
  2-12 Hoshi-ga-oka, Mizusawa-ku, Oshu-shi, Iwate 023-0861}
\altaffiltext{7}{Department of Astronomy, Yonsei University, \\
  134 Shinchong-dong, Seodaemun-gu, Seoul 120-749, Republic of Korea}
\altaffiltext{8}{Department of Astronomy, Graduate School of Science, The University of Tokyo, \\
  7-3-1 Hongo, Bunkyo-ku, Tokyo 113-0033}
\altaffiltext{9}{Korean VLBI Network, Korea Astronomy and Space Science Institute, \\
  P.O.Box 88, Yonsei University, 134 Shinchon-dong, Seodaemun-gu, Seoul 120-749, Republic of Korea}
\altaffiltext{10}{Department of Space Survey and Information Technology, Ajou University, \\
  Suwon 443-749, Republic of Korea}
\email{tomoya.hirota@nao.ac.jp}

\maketitle

\begin{abstract}
We present results of multi-epoch VLBI observations with VERA (VLBI Exploration of 
Radio Astrometry) of the 22~GHz H$_{2}$O masers associated with a young stellar object (YSO) 
IRAS~22198+6336 in a dark cloud L1204G. Based on the phase-referencing VLBI astrometry, 
we derive an annual parallax of IRAS~22198+6336 to be 1.309$\pm$0.047~mas, corresponding 
to the distance of 764$\pm$27~pc from the Sun. 
Although the most principal error source 
of our astrometry is attributed to the internal structure of the maser spots, 
we successfully reduce the errors in the derived annual parallax by employing 
the position measurements for all of the 26 detected maser spots. 
Based on this result, 
we reanalyze the spectral energy distribution (SED) of IRAS~22198+6336 and find that 
the bolometric luminosity and total mass of IRAS~22198+6336 are 
450$L_{\odot}$ and 7$M_{\odot}$, respectively. 
These values are consistent with an intermediate-mass YSO 
deeply embedded in the dense dust core, which 
has been proposed to 
be an intermediate-mass counterpart of a low-mass Class~0 source. 
In addition, we obtain absolute proper motions of the H$_{2}$O masers for the most 
blue-shifted components. 
We propose that the collimated jets aligned along the east-west direction 
are the most plausible explanation for the origin of the detected maser features. 
\end{abstract}

\section{Introduction}

One of the crucial issues in astronomy and astrophysics is to understand 
formation processes of stars, planetary systems, and their clusters. 
They are the most basic constituents in the Galaxy and hence, 
play essential roles in its formation and evolution. 
In order to understand physics and dynamics in the star-formation processes, 
it is necessary to obtain accurate physical properties of target sources 
such as size, mass, and luminosity. 
Because all of these values strongly depend on the adopted distance, 
the distance is the most fundamental parameter for quantitative discussion. 

In spite of its importance, however, it has long been difficult to measure 
accurate distances to star-forming regions without assumptions. 
The most reliable and direct way to determine the distance is an annual trigonometric 
parallax method, based on precise measurements of the position and motion of the object. 
In 1990's, the Hipparcos satellite measured annual parallaxes 
for more than 100~000 stars with a typical precision of 1~mas level 
\citep{perryman1997}. Nevertheless, accurate distance measurements 
with Hipparcos for star-forming regions were limited to nearby sources 
within a few hundred pc from the Sun (e.g. Lombardi et al. 2008). 
Although distances to several OB clusters associated with nearby 
star-forming regions up to $\sim$600~pc have been reported \citep{dezeeuw1999}, 
Hipparcos could not directly determine the distances to 
newly born stars deeply embedded in dusty molecular cloud cores, 
because Hipparcos was capable of observing only in the optical wavelengths. 

In the last few years, phase-referencing VLBI astrometry has been 
developed drastically, 
yielding the annual parallaxes of nearby star-forming regions. 
For instance, radio continuum sources in the Taurus 
\citep{loinard2005, loinard2007, torres2007}, Ophiuchus 
\citep{loinard2008}, and Orion \citep{sandstrom2007, menten2007} 
regions were observed with the NRAO Very Long Baseline Array (VLBA) 
to derive their annual parallaxes with typical uncertainty of 
better than a few percent. 

In addition, we have started the initial scientific project 
"Measurements of annual parallaxes of nearby molecular clouds" 
with VERA (VLBI Exploration of Radio Astrometry). 
VERA is a Japanese VLBI network dedicated to 
astrometric observations aimed at revealing 3-dimensional 
structure of the Galaxy \citep{honma2007, sato2007, sato2008}. 
In our project for nearby molecular clouds, 
we specially focus on the accurate distance measurements of 
star-forming regions within 1~kpc from the Sun \citep{dame1987}. 
Our results will contribute to refine all of the observational 
studies on nearby star-forming regions 
by providing their most accurate and directly measured distances. 
We have already conducted astrometric observations of the H$_{2}$O maser 
sources in the Orion \citep{hirota2007}, Ophiuchus \citep{imai2007}, and 
Perseus \citep{hirota2008} regions. All of the results provide direct measurements 
of their distances with much higher accuracy than the previous photometric methods 
and the parallax measurements with Hipparcos. 

As a part of this project, we present results of astrometric observations 
of the H$_{2}$O maser source IRAS~22198+6336 in a nearby dark cloud core L1204G\footnotemark. 
\footnotetext{In some literature, IRAS~22198+6336 is called as 
L1204A \citep{fukui1989, palla1991, migenes1999}. However, it is 
quite confusing because L1204A is identified to be another 
core associated with the well known massive young stellar object (YSO) 
S140~IRS1 \citep{tafalla1993, tofani1995}. 
Thus, we refer to the host of IRAS~22198+6336 as L1204G in this paper according to 
\citet{tafalla1993}.} 
L1204 is located in the Cepheus-Cassiopeia molecular cloud complex 
\citep{yonekura1997} and is associated with a well known high-mass star-forming 
region S140. It is also identified to be a molecular cloud WB176 \citep{wouterloot1989}. 
According to the molecular line observations of L1204, 
it consists of several high density molecular cores labeled by A-G 
\citep{tafalla1993}. A deeply embedded YSO IRAS~22198+6336 is 
associated with one of the dense cores, L1204G, and it is identified to be a 
powering source of molecular outflow \citep{fukui1989, wilking1989}. 
The centimeter and millimeter counterpart of IRAS~22198+6336 is detected 
with the VLA \citep{sanchez2008}. 
The H$_{2}$O masers were detected \citep{palla1991} and their 
distributions were investigated by 
the VLA \citep{tofani1995} and VLBA \citep{migenes1999}. 
However, they did not measure the proper motions of the H$_{2}$O 
masers because observations by \citet{tofani1995} could not 
achieve sufficient high resolution and \citet{migenes1999} observed 
only at 1 epoch. Therefore, the kinematics of the circumstellar 
materials around IRAS~22198+6336 remains still unclear. 

The distance to L1204 is estimated by a photometric method 
to be 910~pc \citep{crampton1974} assuming that a B0V star HD211880 is 
the ionization source of the H{\sc{ii}} region S140, which is also associated with L1204. 
On the other hand, \citet{dezeeuw1999} identified HD211880 (HIP110125) as a member of 
the Cepheus~OB2 cluster. Using the Hipparcos data, \citet{dezeeuw1999} determined the 
distance to the Cepheus~OB2 cluster to be closer value of 615$\pm$35~pc from the Sun. 
The kinematic distance of 1.3~kpc is sometimes adopted in several literatures 
\citep{wouterloot1989, molinari1996} according to the systemic velocity of $-11$~km~s$^{-1}$ 
\citep{tafalla1993}. Depending on the adopted distance, 
the luminosity of IRAS~22198+6336 could differ by a factor of 4, and hence, 
its physical properties sometimes contain large uncertainties. 

Based on our accurate astrometric observations with VERA, 
we will provide the parallactic distance to IRAS~22198+6336 in L1204G, 
together with the kinematics of the circumstellar gas 
traced by the H$_{2}$O masers. 
We will also discuss the basic properties of the powering source 
of the H$_{2}$O masers by adopting our new distance to L1204G. 

\section{Observations and Data Analysis}

The VERA observations of the H$_{2}$O masers ($6_{1 6}$-$5_{2 3}$, 22235.080 MHz) 
associated with IRAS~22198+6336 have been carried out since 
Nov. 2006 at intervals of about 1~month and are still ongoing. 
We present the results 
from a total of 9 observing sessions 
from Nov. 2006 to Dec. 2007 
(2006/312, 2006/361, 2007/047, 2007/096, 2007/130, 2007/215, 2007/282, 
2007/322, and 2007/353; hereafter an observing session is denoted 
by year/day of the year). 
All 4 VERA stations participated 
in all sessions, providing a maximum baseline length of 2270~km. 

\begin{figure}[tbh]
  \begin{center}
  \FigureFile(80mm,80mm){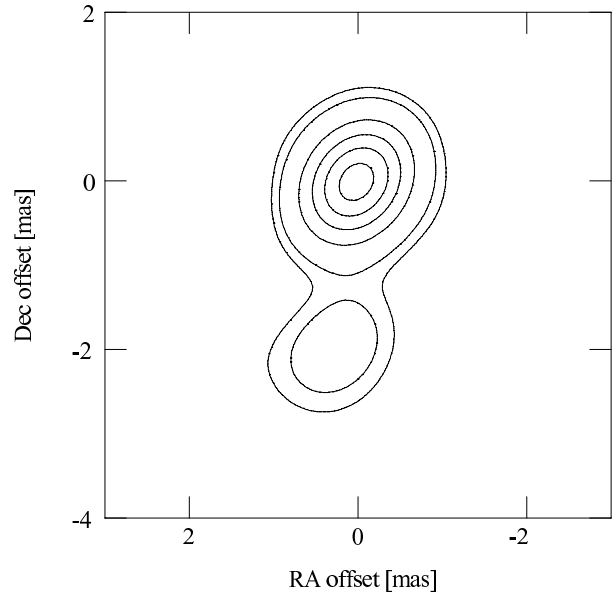}
  \caption{Self-calibrated image of the reference source J2223+6249 
  at the first epoch, 2006/312. 
  The contour levels are 5, 10, 30, 50, 70, and 90\% of the peak intensity
  (205.3~mJy~beam$^{-1}$).}
  \label{fig-j2223}
  \end{center}
\end{figure}

Observations were made in the dual beam mode; the H$_{2}$O masers associated 
with IRAS~22198+6336 and a reference source J2223+6249 
($\alpha_{\rm{J2000.0}}=$22h23m18.096661s, 
$\delta_{\rm{J2000.0}}=+62$d49'33.80393"; \cite{petrov2005}) 
were observed simultaneously \citep{kawaguchi2000, honma2003, honma2008a}. 
The reference source J2223+6249 is separated south from IRAS~22198+6336 by 1.1~degrees. 
J2223+6249 was detected with the peak flux density of about 200~mJy in all the epochs 
as shown in Figure \ref{fig-j2223}. 
Left-handed circular polarization was received and sampled with 2-bit 
quantization, and filtered using the VERA digital filter unit \citep{iguchi2005}. 
The data were recorded onto magnetic tapes at a rate of 1024~Mbps, 
providing a total bandwidth of 256~MHz in which one IF channel and the rest 
of 15 IF channels with 16~MHz bandwidth each were assigned to 
IRAS~22198+6336 and J2223+6249, respectively. 
A bright continuum source, 3C454.3, was observed every 80 minutes 
for bandpass and delay calibration. 
System temperatures including atmospheric attenuation were measured with 
the chopper-wheel method \citep{ulich1976} to be 100-500~K, 
depending on weather conditions and elevation angle of the observed sources. 
The aperture efficiencies of the antennas ranged from 45 to 52\% depending on 
the stations. 
A variation of the aperture efficiency of each antenna as a function of elevation angle 
was confirmed to be less than 10\% even at the lowest elevation in the observations 
($\sim$20~degrees). 
Correlation processing was carried out on the Mitaka FX correlator 
\citep{chikada1991} located at the NAOJ Mitaka campus. 
For H$_{2}$O maser lines, the spectral resolution was set to be 15.625~kHz, 
corresponding to the velocity resolution of 0.21~km~s$^{-1}$. 

Data reduction was performed using 
the NRAO Astronomical Image Processing System (AIPS). 
Amplitude and bandpass calibrations were done for the H$_{2}$O maser source 
(IRAS~22198+6336) and reference source (J2223+6249) independently. 
For phase-referencing, fringe fitting was done with the AIPS task FRING on 
the phase reference source (J2223+6249), and 
the solutions were applied to the target source (IRAS~22198+6336). 
Because the reference source J2223+6249 showed the double-peaked structure 
in the synthesized images 
as shown in Figure \ref{fig-j2223}, 
we carried out self-calibration to correct for 
the structure effect on the phase calibration. 
In addition, we applied the results of the dual-beam phase 
calibration, 
in which difference between instrumentally added phases in 
the two beams were measured 
by injecting artificial noise sources into both beams at each station 
during the observations \citep{kawaguchi2000, honma2008a}. 
We also corrected for the approximate delay model adopted in 
the correlation processing. 
The difference in the optical path length error 
between our newly applied model and the CALC9 model developed by 
the NASA/GSFC VLBI group are confirmed to be less than $\sim$2~mm. 
In this correction, drifts of the visibility phase caused by the Earth's 
troposphere were estimated based on the GPS measurements \citep{honma2008b}. 
As discussed in \citet{honma2008b}, the tropospheric zenith 
delay consists of "dry" and "wet" parts, and only latter one is variable and hence 
significantly cause the phase errors in the calibrations. The typical tropospheric 
"wet" zenith delay (path length) was 5-10~cm while it was sometimes as large as 40 cm, 
in particular at the Ogasawara and Ishigaki stations in summer. On the other hand, 
the "dry" component of the tropospheric zenith delay was stable about 2.3~m. 

Synthesis imaging and deconvolution (CLEAN) were performed using the AIPS task IMAGR 
with the uniform weighting to achieve the highest spatial resolution, allowing us 
to resolve complex spatial structures of the masers. 
The resultant synthesized beam size (FWMH) was typically 
1.2~mas$\times$0.9~mas with a position angle of -40~degrees. 
The peak positions and peak flux densities 
of masers were derived by fitting elliptical Gaussian brightness distributions 
to each image using the AIPS task SAD. 
The formal uncertainties in the maser positions given by SAD were 
mostly better than 0.05~mas. 

\section{Results}
\subsection{Overall Properties of the H$_{2}$O Masers in IRAS~22198+6336}
\label{sec-distribution}

\begin{figure}[tbh]
  \begin{center}
    \FigureFile(80mm,80mm){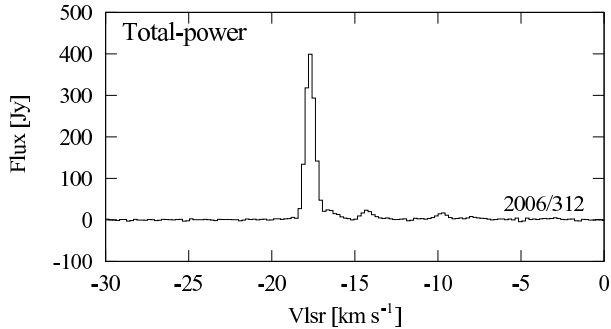}
  \caption{Total-power spectrum of the H$_{2}$O masers 
  associated with IRAS~22198+6336 observed at the first epoch, 2006/312. 
  The spectrum is the average of those observed at all the VERA stations. }
  \label{fig-total}
  \end{center}
\end{figure}

\begin{figure*}[tbh]
  \begin{center}
    \FigureFile(140mm,140mm){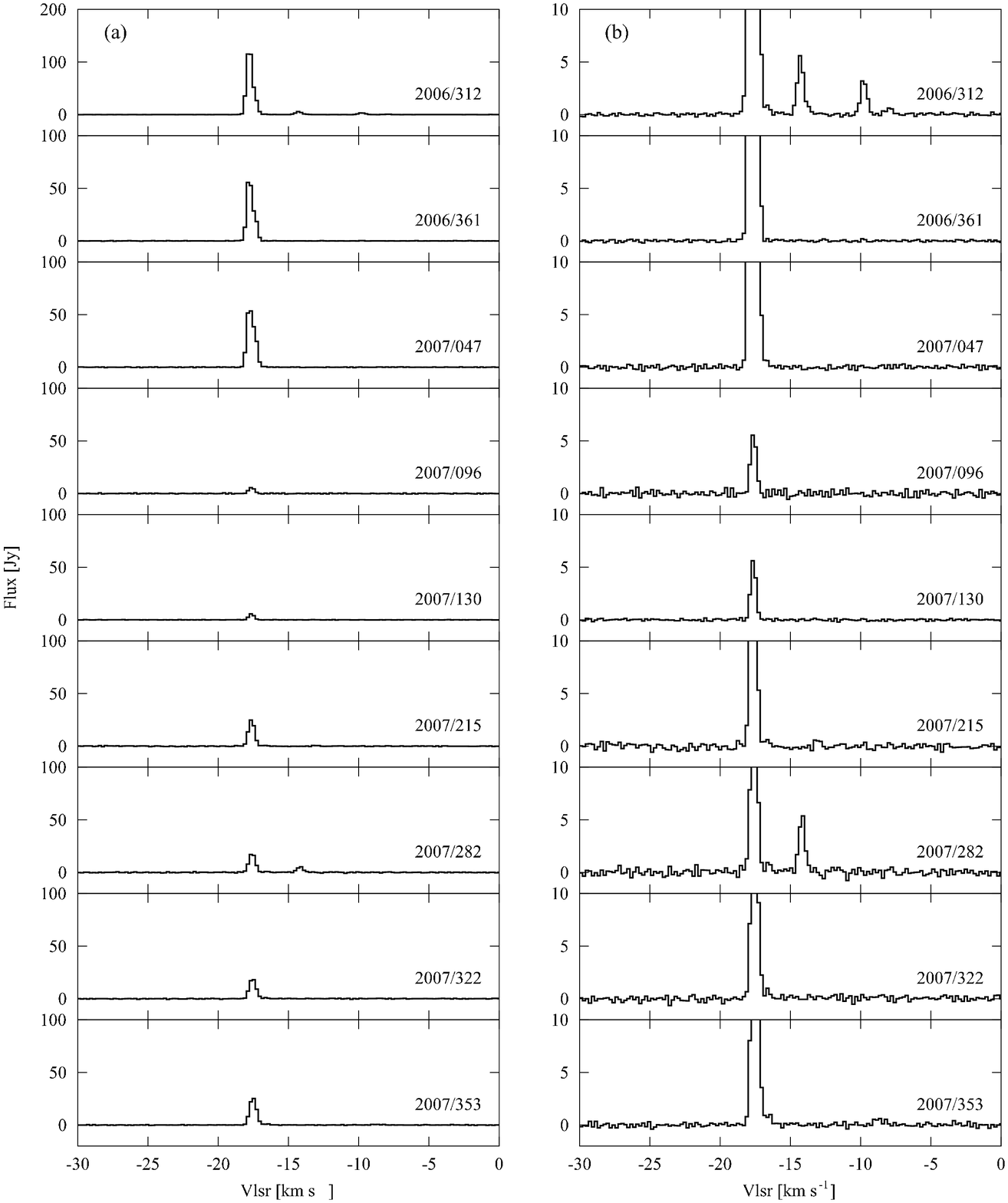}
  \caption{(a) Scalar-averaged cross-power spectra of the H$_{2}$O masers 
  associated with IRAS~22198+6336. 
  The spectra are the averages of those observed at all the VERA stations. 
  Note that the scale of the ordinate for 
  the 2006/312 epoch is twice as large as others. (b) Same as (a) but 
  magnified by a factor of 10 to clarify the weaker spectral components. }
  \label{fig-spectra}
  \end{center}
\end{figure*}

Figures \ref{fig-total} and \ref{fig-spectra} show an example of 
a total-power spectrum and scalar-averaged cross-power spectra of 
the H$_{2}$O masers associated with IRAS~22198+6336, respectively. 
The H$_{2}$O maser lines were detected 
within the LSR (local standard of rest) velocity range from $-18.5$~km~s$^{-1}$ to $-2.7$~km~s$^{-1}$. 
The intense spectra at the LSR velocity of about $-17$~km~s$^{-1}$ 
were detected in all the observed epochs, whereas 
other velocity components were detected at rarer occurrences. 
In this paper, we refer to the brightest components at $-17$~km~s$^{-1}$ 
as the main components. 
Our result is consistent with those of long-term monitoring 
observations made by \citet{valdettaro2002} and 
\citet{brand2003} that the most prominent maser feature in IRAS~22198+6336 was at the LSR 
velocity of about $-20$~km~s$^{-1}$, while those around the systemic 
velocity of $-11$~km~s$^{-1}$ \citep{tafalla1993} rarely appeared. 
According to their long-term monitoring observations, 
the radial velocities of the H$_{2}$O masers in IRAS~22198+6336 changed 
with time over the 11-year monitoring period at a rate of 
0.2~km~s$^{-1}$~yr$^{-1}$. 
The LSR velocity of the $-20$~km~s$^{-1}$ component has been gradually 
red-shifted to be about $-18$~km~s$^{-1}$ \citep{valdettaro2002, brand2003}. 
This trend predicts $\sim -17$~km~s$^{-1}$ of the velocity of the main component 
at around 2007, as observed in our observations. 
However, we cannot find any signature of the velocity drift 
within our observing period probably due to our relatively shorter 
time span. 

Since the spectrum taken at the first epoch, 2006/312, showed 
most of the velocity components, 
we first searched for the H$_{2}$O masers at the first epoch 
by producing the synthesized 
images for all the individual velocity channels with the field of view of 
$\sim$4\arcsec$\times$4\arcsec. 
As a result, we found roughly 4 clusters of maser spots, or features, spreading over 
the 300~mas$\times$300~mas region. 
Hereafter we define a ``spot" as emission occurring in a single velocity channel and 
a ``feature" as a group of spots which are spatially coincident within the beam 
size and are detected for at least two consecutive velocity channels. 
Based on this coarse map, we again carried out synthesis imaging for all the epochs 
with the field of view of 25.6~mas$\times$25.6~mas 
centered at the positions for each of the identified features. 

\begin{figure*}[tbh]
  \begin{center}
    \FigureFile(160mm,160mm){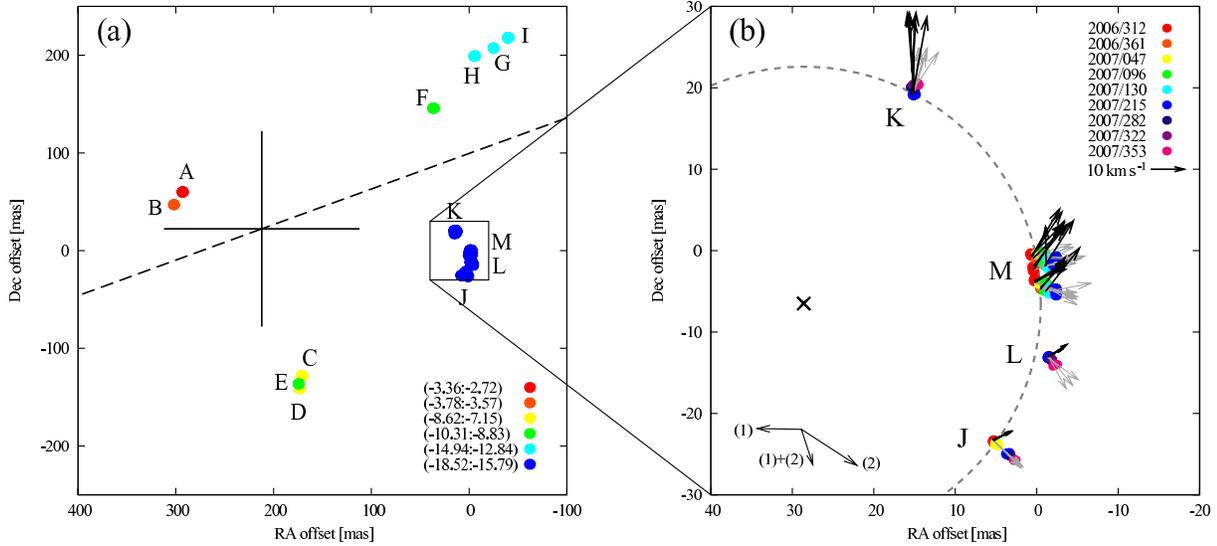}
  \caption{Phase-referenced maps of the H$_{2}$O masers associated with IRAS~22198+6336. 
   (a) Overall distribution of the H$_{2}$O maser spots. 
   All the detected spots for all of the 9 observing epochs are plotted. 
   The position offsets in right ascension and declination are 
   measured with respect to the (0,0) position at 
   $\alpha(J2000.0)=$22h21m26.72790s, $\delta(J2000.0)=+63^{\circ}$51\arcmin37.9239\arcsec. 
   A cross and a dashed line indicate the peak position and the direction of its elongation 
   of the 7~mm radio continuum emission \citep{sanchez2008}, of which error bar is 
   roughly estimated from the beam size and signal-to-noise ratio of the VLA observation. 
   (b) Distribution and proper motions of the maser spots in the main 
   features J-M. 
   The movements of the spots indicated by the color-coded filled circles represent the absolute 
   proper motions without correction of the Solar motion and Galactic rotation. 
   Grey arrows represent the absolute proper motions whereas the thick black arrows 
   represent the proper motions corrected for the Solar motion and the Galactic rotation. 
   The proper motion of 1~mas~yr$^{-1}$ corresponds to 3.6~km~s$^{-1}$, 
   and the proper motion vector of 10~km~s$^{-1}$ is shown in the top left corner of this figure. 
   The arrows shown in the bottom-left corner represent the magnitude of  
   (1) the Solar motion (3.612~mas~yr$^{-1}$, $0.064$~mas~yr$^{-1}$), 
   (2) the Galactic rotation ($-4.543$~mas~yr$^{-1}$, $-2.947$~mas~yr$^{-1}$), 
   and the sum of (1) and (2). 
   A large circle indicates the best fit model of the possible expanding shell 
   model (see text). }
  \label{fig-allspot}
  \end{center}
\end{figure*}

\begin{figure}[tbh]
  \begin{center}
    \FigureFile(80mm,80mm){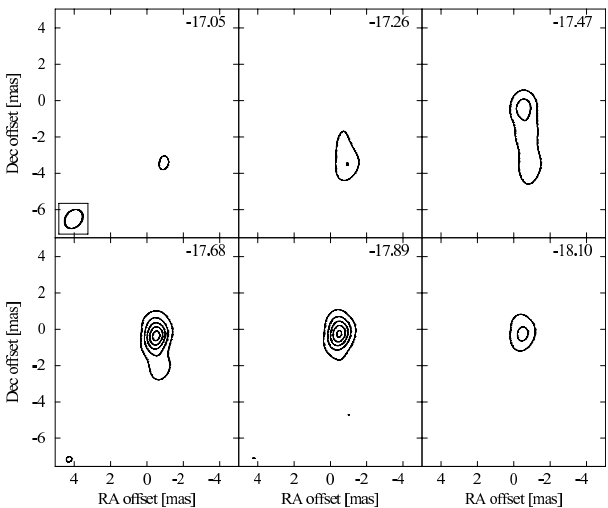}
  \caption{Examples of phase-referenced channel maps of the elongated maser features M 
  at the first epoch, 2006/312. 
  Contour levels are 10, 30, 50, 70, and 90\% of the peak intensity (112.2~Jy~beam$^{-1}$). 
  The LSR velocities are indicated at the top-right corner of each panel, in unit of km~s$^{-1}$. 
  The beam size is shown in the top-left panel. }
  \label{fig-chmap}
  \end{center}
\end{figure}

The results of the mapping are shown in Figure \ref{fig-allspot}, 
and examples of phase-referenced channel maps are shown in Figure \ref{fig-chmap}. 
We labeled the groups of features A-M in the order of their LSR velocities. 
Some of the weak components were not detected in the phase-referenced maps 
but only detected in the self-calibrated images, in which we carried out 
phase-calibration using the brightest maser spot rather than the reference 
source J2223+6249, due to the lower dynamic range of the phase-referenced images. 
For such spots, we derived their absolute positions by referencing to 
the absolute position of the brightest spot in the main component that is 
significantly detected in the phase-referenced images at all the epochs. 

The maser distribution shows roughly a circular structure with the radius of about 150~mas 
as seen in Figure \ref{fig-allspot}(a). 
Overall spatial and radial velocity distribution of H$_{2}$O maser features are in 
good agreement with those by the VLA \citep{tofani1995} and the VLBA 
\citep{migenes1999}, but we resolve the previously detected components 
into several spatially distinct maser spots due to our higher spatial resolution. 
The main components which are most blue-shifted and brightest, J-M, coincide with 
both the position and velocity of 
the C2 component labeled by \citet{tofani1995}, 
while the most red-shifted components, A-B, coincide 
in the position and velocity 
with the C4 component of \citet{tofani1995}. 
The northern components near the systemic velocity, F-I, are also coincident 
with both the position and velocity of the C1 component of \citet{tofani1995}. 
On the other hand, another group of components C-E is spatially coincident with 
C3 of \citet{tofani1995}, whereas their LSR velocities, from -7 to -10~km~s$^{-1}$, 
are significantly red-shifted by more than 5~km~s$^{-1}$. 
Here we emphasize that the accuracy of our absolute positions with an uncertainty of 
only 1~mas, which is limited mainly by the position uncertainty of 
the reference source J2223+6249, 
0.90~mas and 1.31~mas in right ascension and declination, 
respectively \citep{petrov2005}, 
are about two orders of magnitude higher than those 
by the previous VLA observations \citep{tofani1995} and 
by the fringe-rate maps obtained with the VLBA \citep{migenes1999}. 

The properties of the features A-I are summarized in Table \ref{tab-features}. 
These parameters are derived from the phase-referenced images except for the 
two features, D and E, as noted in Table \ref{tab-features}. 
They are compact and almost no significant spatial structures, and hence, 
we only summarize the parameters for the peak spot in each feature. 
All of the features listed in Table \ref{tab-features} were detected 
each only at a single one different epoch 
probably due to their high variability. 

\begin{figure*}[tbh]
  \begin{center}
    \FigureFile(160mm,160mm){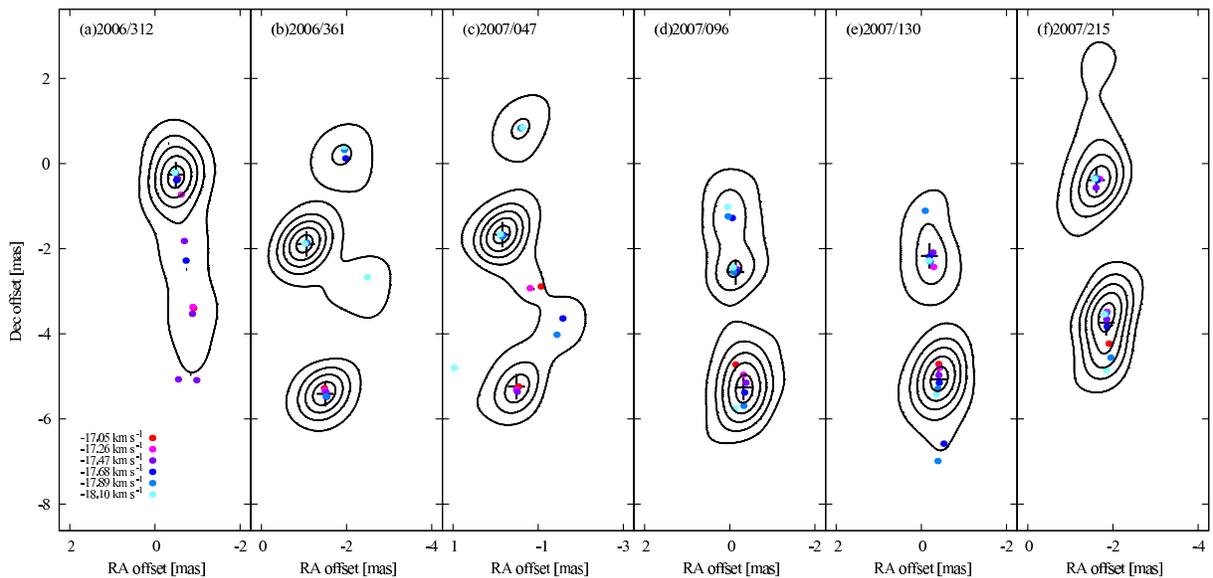}
  \caption{Structures of the elongated maser features M 
   obtained by integrating the phase-referenced channel maps. 
   Contour maps show the integrated flux densities of the features with the contour levels 
   of 10, 30, 50, 70, and 90\% levels of the peak flux densities. 
   Peak values are 329.9~Jy~beam$^{-1}$, 60.0~Jy~beam$^{-1}$, 52.0~Jy~beam$^{-1}$, 
   13.1~Jy~beam$^{-1}$, 26.1~Jy~beam$^{-1}$, and 16.2~Jy~beam$^{-1}$, for 
   (a)2006/312, (b)2006/361, (c)2007/047, (d)2007/096, (e)2007/130, 
   and (f)2007/215, respectively. 
   Crosses represent the positions and their error bars for the features 
   employed in the datasets 3 and 4 (see Table \ref{tab-results}), 
   while color-coded filled circles represent the positions of the spots.}
  \label{fig-central}
  \end{center}
\end{figure*}

On the other hand, the brightest main features, J-M, show complex spatial structures. 
We summarize the parameters for the selected 26 spots in Table \ref{tab-spots}, 
which are detected in at least three epochs, and hence, are usable to 
obtain reliable proper motions as discussed later. 
The parameters for the spots L1-3 are derived from the self-calibrated images while others 
are from the phase-referenced images. 
The main features are resolved into four subgroups; 
J $(4~{\rm{mas}}, -23~{\rm{mas}})$, K $(15~{\rm{mas}}, 20~{\rm{mas}})$, 
L $(-1~{\rm{mas}}, -12~{\rm{mas}})$, and M $(-1~{\rm{mas}}, -3~{\rm{mas}})$ 
in right ascension and declination, respectively, 
and all of them consist of 
multiple spots. In particular, the component M shows the elongated 
structure along the north-south direction from 0~mas to $-6$~mas in declination, 
as shown in Figures \ref{fig-central} and \ref{fig-chmap}. 
The relatively large scale structures found in these features have been inferred 
from the fact that the H$_{2}$O masers are highly 
resolved-out with the longer baselines in the VLBA \citep{migenes1999}. 
We also found that the flux densities of the cross-power spectra 
of the masers are typically only about 30\% of those of the total-power spectra 
as shown in Figures \ref{fig-total} and \ref{fig-spectra}, 
which are suggestive of the missing flux. 
Although the peak velocities of the main features, $-17$~km~s$^{-1}$ were stable, 
their internal structures were found to be highly variable. 
The features K and L appeared later than 2007/215, while 
M was detected only at the earlier epochs than 2007/215. 
In contrast, the feature J was highly variable but detected throughout whole 
of the observing period. 

As depicted in Figure \ref{fig-allspot}(a), 
we found a radial velocity gradient along the east-west direction 
when we consider the clusters of features A-B and J-M; 
blue-shifted (J-M) and red-shifted(A-B) components tend to be distributed 
at the western and eastern sides of the map, respectively. 
The similar radial velocity gradient has been reported by 
\citet{tofani1995} and \citet{migenes1999} which is attributed to 
the bipolar outflow. 
On the other hand, the larger-scale ($>$1\arcmin) velocity gradient along 
the northwest-southeast direction has been found in the spectral 
line observations of NH$_{3}$ \citep{tafalla1993} and C$^{18}$O \citep{jenness1995}. 
Although they attributed these velocity gradients to the molecular outflow, 
their velocity structures showed an opposite trend to our H$_{2}$O maser 
distribution; the blue-shifted and red-shifted components traced by 
the NH$_{3}$ and C$^{18}$O lines tend to be 
located at southeast and northwest, respectively. 
It would be suggested that the large-scale velocity gradient does not 
directly related to that traced by the H$_{2}$O masers. 
Detailed discussion about the spatial and velocity structures of the H$_{2}$O 
masers will be presented later. 

\begin{table*}[tbh]
\begin{center}
\caption{The peak maser spots in the features A-I} 
\label{tab-features}
\begin{tabular}{lccrrr}
\hline
\hline
   & $v_{lsr}$     & Detected & $\Delta \alpha$ & $\Delta \delta$ & $F_{peak}$       \\ 
No & [km s$^{-1}$] & Epoch   & [mas]           & [mas]           & [Jy beam$^{-1}$] \\
\hline
A  &  -3.15 &   2006/312   &   293.43   &     60.22   &   1.9 \\ 
B  &  -3.57 &   2006/312   &   302.30   &     47.19   &   0.74 \\ 
C  &  -8.20 &   2006/312   &   171.42   &   -128.00   &   3.9 \\ 
D$^{a}$ &  -8.41 &   2007/353   &   173.85   &   -141.05   &   2.2 \\ 
E$^{a}$ &  -8.83 &   2007/353   &   174.73   &   -136.19   &   2.7 \\ 
F  &  -9.68 &   2006/312   &    36.88   &    145.81   &  11.5 \\ 
G  & -13.05 &   2007/215   &   -24.59   &    207.43   &   1.9 \\ 
H  & -14.31 &   2006/312   &    -5.22   &    199.21   &  15.7 \\ 
I  & -14.31 &   2007/353   &   -39.40   &    218.13   &   7.2 \\ 
\hline 
\hline 
\multicolumn{6}{l}{Note --- All the features are detected only at the epoch } \\
\multicolumn{6}{l}{noted in the third column. The brightest spots in each } \\
\multicolumn{6}{l}{feature are listed. } \\
\multicolumn{6}{l}{${a}$: The spots are detected only in the self-calibrated }\\
\multicolumn{6}{l}{images. } \\
\end{tabular}
\end{center}
\end{table*}

\begin{table*}[tbh]
\begin{center}
\caption{All the maser spots in the main features J-M used for the parallax measurement} 
\label{tab-spots}
\begin{tabular}{lccrrrrr}
\hline
\hline
   & $v_{lsr}$     & Detected & $\Delta \alpha$ & $\Delta \delta$ & $F_{peak}$       & $\mu_{\alpha} \cos \delta$ & $\mu_{\delta}$  \\ 
No & [km s$^{-1}$] & Epochs   & [mas]           & [mas]           & [Jy beam$^{-1}$] & [mas yr$^{-1}$]            & [mas yr$^{-1}$] \\
\hline
J1 & -16.42 &  1*3**67*9  &    4.15 &    -23.17  &   3.7 &   -2.33(25)  &   -2.32(32)   \\ 
J2 & -16.63 &  1*3**67*9  &    4.14 &    -23.31  &   6.4 &   -2.48(25)  &   -2.20(32)   \\ 
J3 & -16.84 &  1*3**67*9  &    4.03 &    -23.33  &   3.0 &   -2.27(25)  &   -2.10(32)   \\ 
K1 & -17.05 &  ******789  &   14.21 &     20.60  &   2.7 &   -2.02(168) &    3.00(213)  \\ 
K2 & -17.26 &  *****6789  &   15.53 &     20.48  &   4.7 &   -0.99(83)  &    3.18(105)  \\ 
K3 & -17.47 &  *****6789  &   15.51 &     20.44  &   8.2 &   -0.58(83)  &    3.71(105)  \\ 
K4 & -17.68 &  *****6789  &   15.48 &     20.46  &   8.5 &   -0.55(83)  &    3.71(105)  \\ 
K5 & -17.89 &  *****6789  &   15.51 &     20.46  &   6.8 &   -0.40(83)  &    3.79(105)  \\ 
K6 & -18.10 &  *****67*9  &   15.62 &     20.36  &   2.7 &   -1.29(88)  &    3.80(110)  \\ 
L1$^{a}$ & -17.68 &  *****6789  &   -0.99 &    -11.88  &  12.8 &   -1.45(83)  &   -2.63(105)  \\ 
L2$^{a}$ & -17.89 &  *****6789  &   -0.95 &    -11.94  &  10.1 &   -2.53(83)  &   -1.85(105)  \\ 
L3$^{a}$ & -18.10 &  *****6789  &   -1.02 &    -11.77  &   2.5 &   -2.00(83)  &   -2.44(105)  \\ 
M1 & -17.05 &  123456***  &   -0.90 &     -3.40  &  19.6 &   -3.44(41)  &   -2.00(50)   \\ 
M2 & -17.26 &  123456***  &   -0.88 &     -3.37  &  60.2 &   -3.45(41)  &   -1.23(50)   \\ 
M3 & -17.47 &  123456***  &   -0.87 &     -3.53  &  48.8 &   -3.43(41)  &   -1.35(50)   \\ 
M4 & -17.68 &  *2*456***  &   -1.52 &     -5.48  &  10.4 &   -2.98(55)  &   -0.75(69)   \\ 
M5 & -17.89 &  *2*456***  &   -1.51 &     -5.47  &   8.9 &   -2.96(55)  &   -0.26(69)   \\ 
M6 & -18.10 &  ***456***  &   -0.12 &     -5.75  &   0.6 &   -3.73(98)  &    0.70(125)  \\ 
M7 & -17.47 &  1**456***  &   -0.68 &     -1.82  &  42.6 &   -3.49(45)  &    0.26(55)   \\ 
M8 & -17.68 &  1**456***  &   -0.72 &     -2.28  &  20.4 &   -3.44(45)  &    1.11(55)   \\ 
M9 & -17.89 &  ***456***  &   -0.08 &     -2.55  &   1.2 &   -3.28(98)  &    0.83(125)  \\ 
M10 & -18.10 &  *2*456***  &   -2.48 &     -2.67  &   3.3 &   -0.90(55)  &    0.30(69)   \\ 
M11 & -17.05 &  1*3**6***  &   -0.52 &     -0.35  &   2.7 &   -3.89(45)  &   -0.01(56)   \\ 
M12 & -17.68 &  1234*****  &   -0.50 &     -0.39  & 175.7 &   -3.39(78)  &    0.80(96)   \\ 
M13 & -17.89 &  12345****  &   -0.47 &     -0.22  & 146.1 &   -3.62(60)  &   -0.24(72)   \\ 
M14 & -18.10 &  1234*****  &   -0.45 &     -0.21  &  59.5 &   -3.24(78)  &    0.99(96)   \\ 
\hline
\hline 
\multicolumn{8}{l}{Note --- In the third column, the numbers indicate the detected epochs corresponding to } \\
\multicolumn{8}{l}{1; 2006/312, 2; 2006/361, 3; 2007/047, 4; 2007/096, 5; 2007/130, 6; 2007/215, 7; 2007/282, } \\
\multicolumn{8}{l}{8; 2007/322, and 9; 2007/353, respectively. The epoch in which the maser is not } \\
\multicolumn{8}{l}{detected is indicated by an asterisk instead of above number. Positions and flux densities } \\
\multicolumn{8}{l}{are those observed at the epoch detected for the first time, as noted in the third column. } \\
\multicolumn{8}{l}{Numbers in parenthesis represent the errors in unit of the last significant digits. } \\
\multicolumn{8}{l}{${a}$: In the epochs of 2007/322 and 2007/353, the spots are detected only in the } \\
\multicolumn{8}{l}{self-calibrated images.} \\
\end{tabular}
\end{center}
\end{table*}

\subsection{Astrometry of the H$_{2}$O Masers}

We conducted monitoring observations of 
the H$_{2}$O masers for about one year, and 
the absolute motions of the maser spots were successfully 
obtained by referencing to the extragalactic 
position reference source J2223+6249. 
The movement of each maser spot can be expressed by the sum of a 
linear motion and the annual parallax. 
It can be assumed that all the maser features or spots in L1204G 
have the same annual parallax within the astrometric accuracy of VERA. 
Therefore, we can simultaneously determine the proper motions 
($\mu_{\alpha}^{i} \cos \delta$, $\mu_{\delta}^{i}$) and 
the initial positions ($\alpha_{0}^{i}$, $\delta_{0}^{i}$) in 
right ascension and declination for the $i$-th feature or spot, 
along with a common annual parallax of 
L1204G, $\pi$, by the least-squares analysis of 
all the detected features or spots. 
Because only the features in the main components, J-M, were detected 
at more than two epochs, we excluded the features A-I in 
the following astrometric analysis. 

To obtain the accurate annual parallax, we attempted to analyze the 
data by adopting four different datasets; 
(1) both the right ascension and declination data for all the spots, 
(2) only the right ascension data for all the spot, 
(3) same as (1) but for all the features, and 
(4) same as (2) but for all the features. 
As already defined, 
the feature is the sum of the spatially coincident 
maser spots integrated over at least two consecutive velocity channels. 
Thus, we determined the peak position of each feature by the Gaussian fitting 
of the velocity-integrated phase-referenced intensity map. 
On the other hand, the peak position of 
each spot is derived from the individual channel map obtained from 
the phase-referenced analysis by assuming that 
the velocity of the maser spot is constant, 
i.e. the maser spots are regarded to be identical if their velocities are the same. 
It would not always be valid if the masers show significant velocity drift 
\citep{hirota2008}. However, we applied this assumption to the identification of 
the detected spots between different epochs because we could not find 
significant velocity drift in our observations as mentioned above. 

\begin{figure*}[tbh]
  \begin{center}
    \FigureFile(140mm,140mm){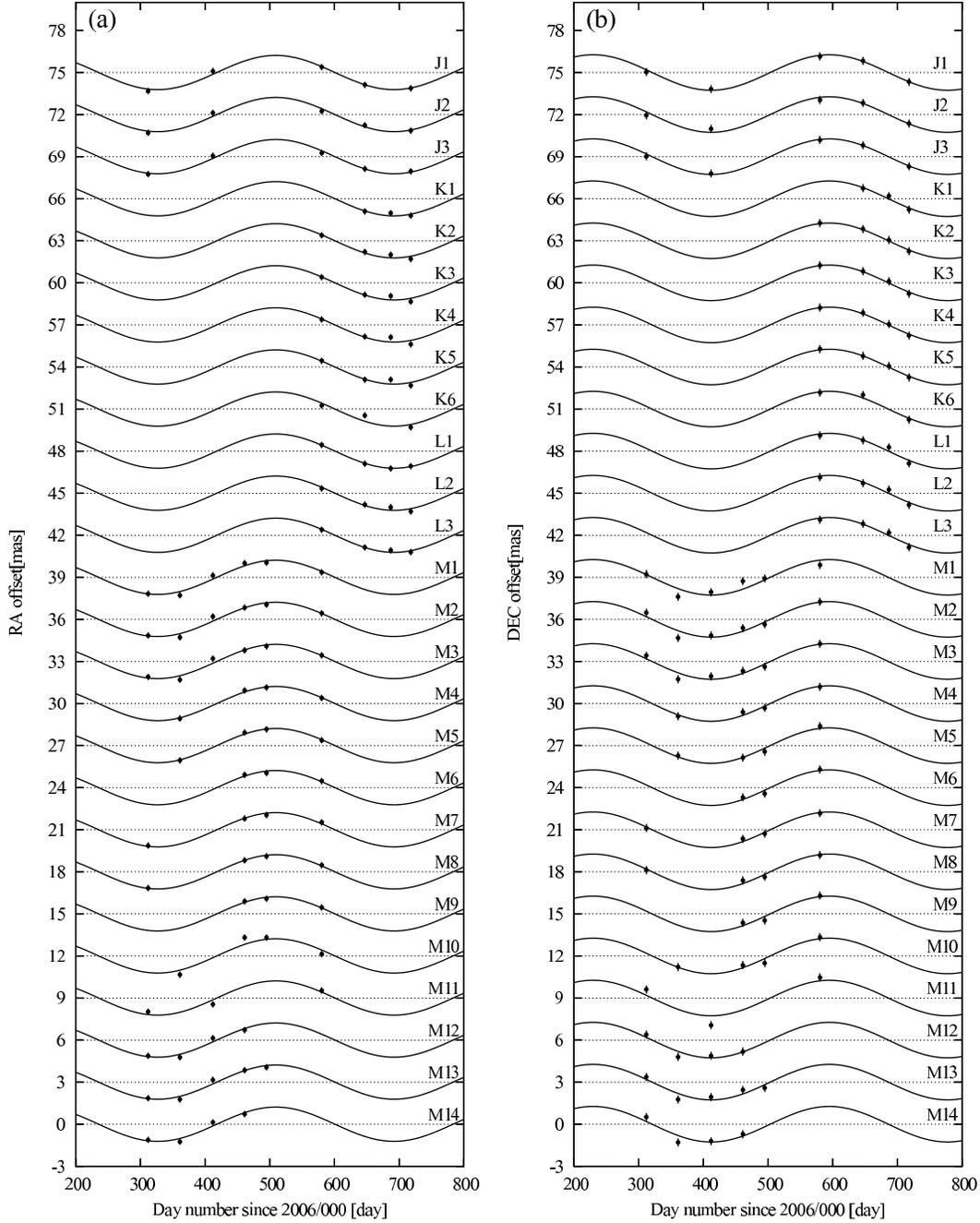}
  \caption{Position measurements of the maser spots in the features J-M. 
   (a) The movement in right ascension as a function of time. 
   Solid lines represent the best fit model of the annual parallax which are 
   common for all the spots. 
   The best fit proper motions are removed, and hence, the curves represent the 
   modulation in the positions due to the annual parallax. 
   Filled circles represent the observed positions of the maser spots. 
   (b) The same as (a) in declination. The associated error bars, 0.23~mas and 0.29~mas 
   in right ascension and declination, respectively, are also plotted but cannot be seen 
   in the figures. }
  \label{fig-parallax}
  \end{center}
\end{figure*}

Comparison of former and latter datasets shows the effect of the 
structures in the maser features on our astrometric accuracy. 
As mentioned above, we found significant spatial structures in the features J-M. 
In particular, the elongated structure in the feature M 
can be clearly seen in Figure \ref{fig-central}, 
in which the peak positions of the maser spots do not always coincide 
with those of the features. 
These complex spatial and velocity structures within the feature M 
would degrade the accuracy of astrometry as reported in 
the previous results of VERA for nearby sources 
\citep{hirota2007, hirota2008, imai2007}. 
Instead, we resolved the internal structure 
of the feature M by producing the channel maps of all the velocity components. 
They could better trace the motion of the masing gas in the multi-epoch astrometric analysis. 
According to their positions and velocity channels, we identified in total 
5 features (1 in feature J-L 
for each and 2 in feature M) and 
26 spots in the main components J-M as listed in Table \ref{tab-spots}, 
which were detected in at least three observing epochs. 
Some of the spots, for instance, the spots J1-3, K6, M4-5, M7-8, and M10-11, 
were not always detected probably due to the variability 
and the insufficient sensitivity. 
However, our identification of the maser spots is proved to be appropriate 
because their motions are well fitted to the linear proper motions and 
annual parallax which is common for other spots, as shown in Figure \ref{fig-parallax}. 
Thus, all of the proper motions identified in our analysis really 
trace the bulk motion of the masing gas. 

In addition, we evaluate the effect of the atmospheric zenith delay 
residual by comparing the datasets 1 and 3 with those of datasets 2 and 4. 
According to the previous results with VERA 
\citep{honma2007, hirota2007, hirota2008, imai2007}, 
the accuracy of the derived annual parallax is significantly improved 
when only the data for right ascension are used, as this data is less 
affected by atmospheric modeling errors. 
Therefore, we ran the fitting by making use of right ascension data alone 
and both right ascension and declination data for comparison. 

All the results for different four datasets can be compared in Table \ref{tab-results}. 
The results of the fitting are also displayed in Figure \ref{fig-parallax} for the dataset 1. 
The standard deviations of the post-fit residuals are $\sim$0.2~mas and $\sim$0.3~mas 
in right ascension and declination, respectively. 
If we employ the formal uncertainties in the Gaussian fitting 
of the maser spots, $\sim$0.05, as the positional errors in our observations, 
a reduced-$\chi^{2}$ is as large as 20. 
This result would suggest that we underestimate the positional errors and 
that these post-fit residuals represent the actual positional accuracy of 
the astrometric observations with VERA. 
Thus, we took into account $\sigma_{\alpha}$ and $\sigma_{\delta}$ listed in 
Table \ref{tab-results} in the least-squares analysis to give a weight, 
which is inversely proportional to the square of the error, for the 
right ascension and declination data, respectively. 

As can be seen in Table \ref{tab-results}, all of the 4 datasets yield 
the consistent parallax values within the mutual errors. 
Although none of the spots were persistent during whole of the observing 
period due to their variability, 
we could determine the annual parallax with an uncertainty of smaller than 10\% 
by successfully identifying multiple spots during almost 1~year. 
The highest accuracy is achieved by adopting both right ascension and declination data 
for all of the 26 detected spots, dataset 1. 
The resultant value of the annual parallax for the dataset 1 is 
1.309$\pm$0.047~mas, corresponding the distance to IRAS~22198+6336 of 764$\pm$27~pc. 
The associated uncertainties of only $\sim$4\% for the dataset 1 is smaller than 
the others, and this dataset provides the most reliable value. 
Detailed discussions about the possible astrometric 
error sources will be presented in the section \ref{sec-error}. 

\begin{table*}[tbh]
\begin{center}
\caption{Summary of the least-squares analysis for the annual parallax measurements}
\label{tab-results}
\begin{tabular}{lcccc}
 \hline\hline
Parameter                         & Dataset 1     & Dataset 2    & Dataset 3    & Dataset 4    \\ 
\hline
Dataset                           & RA and Dec    & RA only      & RA and Dec   & RA only      \\
                                  & 26 spots      & 26 spots     & 5 features   & 5 features   \\
Total number of data              &       218     &      109     &      52      &     26       \\
Numbers of fitted parameters      &       105     &       53     &      21      &     11       \\
Degrees of freedom                &       113     &       56     &      31      &     15       \\
$\pi$ (mas)                       &     1.309(47) &    1.289(66) &    1.243(80) &  1.288(107)  \\
$D$ (pc)                          &       764(27) &      776(40) &      805(52) &     777(65)  \\
$\sigma_{\alpha}$ (mas)           &       0.23    &      0.23    &      0.22    &     0.23     \\
$\sigma_{\delta}$ (mas)           &       0.29    & ---          &      0.32    & ---          \\
\hline \\
\multicolumn{5}{l}{Note --- Numbers in parenthesis represent the errors in unit of the } \\
\multicolumn{5}{l}{last significant digits. $\sigma_{\alpha}$ and $\sigma_{\delta}$ are 
the rms deviations of the post-fit residuals } \\
\multicolumn{5}{l}{in the right ascension and declination directions, respectively. } \\
\end{tabular}
\end{center}
\end{table*}

Along with the annual parallax, proper motions of the maser spots are derived, 
as summarized in Table \ref{tab-spots} and illustrated 
in Figure \ref{fig-allspot}(b). The proper motion of 1~mas~yr$^{-1}$ corresponds 
to 3.6~km~s$^{-1}$ at the distance of 764~pc. 
Although the proper motions for some of the features contain large uncertainties, 
we can see a systematic trend in the direction and magnitude of the proper motion 
vector in Figure \ref{fig-allspot}(b). The kinematics traced by the proper motions 
as well as the radial velocities of the masers will be discussed in the 
section \ref{sec-model}. 

\section{Discussions}

\subsection{Astrometric Error Sources}
\label{sec-error}

As discussed in the recent works on astrometry with VLBA and VERA 
\citep{hachisuka2006, honma2007}, 
it is difficult to estimate individual error sources induced in the VLBI astrometry 
quantitatively. 
In this paper, we regard the post-fit residuals of the least-squares analysis 
listed in Table \ref{tab-results} as the typical positional errors, which 
results in a reduced-$\chi^{2}$ of unity. 
Because they are significantly larger than the formal uncertainties 
in the Gaussian fitting of the maser spots, $\sim$0.05~mas, some systematic errors 
would have significant effects on the astrometry. 
Here we discuss about the possible error sources in the present results. 

One of the serious error sources is the atmospheric zenith delay residuals mainly 
due to the tropospheric water vapor. 
This effect is caused by the difference in the optical path lengths through the atmosphere 
between the target and reference sources because their elevation angles are usually different. 
\citet{honma2008b} presented results of detailed simulations to calculate 
the positional errors caused by the atmospheric zenith delay residuals in 
astrometry with VERA. They demonstrated that 
the positional error due to the zenith delay residual of 2~cm, 
which is a typical value of VERA observations, is better than 0.030~mas 
and 0.018~mas for right ascension and declination, respectively, 
in the case of the declination of 60~degrees, 
separation angle (SA) between the two sources of 1~degree, 
and the position angle (PA) on the sky from the target source to the reference 
source of 180~degrees, which are valid for the IRAS~22198+6336 and J2223+6249 
pair aligned along the north-south direction. 
It turns out that the simulated positional error for this pair 
cannot fully explain the observed positional uncertainties, unless we introduce an 
unrealistically large scatter of the zenith delay residual of $\sim$30~cm for 
each of the observation. 
Therefore, it is unlikely that the atmospheric zenith 
delay residuals is the most dominant source of our astrometric errors. 

Alternatively, we propose that the principal error source in the present 
results would be the temporal variation of the internal structure of the masers,
as proposed by some previous results from the VERA astrometry 
\citep{hirota2007, hirota2008, imai2007}. 
This effect would be more serious for nearby sources than the distant ones. 
In fact, we can see spatial structure of the features in Figure \ref{fig-central}, 
and some of the spots in Figure \ref{fig-parallax} show significant 
deviations from the best fit model. 
Even in the case of the fitting of the spots rather than the features, 
the internal structures in the individual spots might be still significant. 
Although this effect would limit 
the positional accuracy of each observation, the accuracy of the annual 
parallax measurements can be improved with increasing number of spots used in the analysis 
if the variation of the structure behaves in the random way. 
Indeed, the post-fit residuals are reduced almost inversely proportional to the square 
root of the degrees of freedom in the fitting as listed in Table \ref{tab-results}. 

On the other hand, we confirmed that the uncertainties in the 
station positions, path length errors due to ionosphere, and the delay model employed 
in our data analysis, which is consistent with the CALC9 model, would 
have negligible effects on our astrometry with VERA \citep{honma2007}. 
The uncertainties in the absolute position of the reference source J2223+6249, 
0.90~mas and 1.31~mas in right ascension and declination, respectively 
\citep{petrov2005}, do not affect the derived annual parallax and 
proper motion because they would add only a constant offset to the position of 
the maser features. The spatial structure and its temporal variation of 
the reference source J2223+6249 possibly contribute to the positional errors of 
the maser features. However the structure effect of the reference source 
would be much less significant than those of the masers because we performed 
self-calibration of J2223+6249 to solve its structure. According to the dynamic range 
of the self-calibrated image of J2223+6249 ($\sim$20; Figure \ref{fig-j2223}), 
the positional uncertainty caused by the structure of the reference source 
would be less than 0.025~mas (HPBW/2SNR where HPBW and SNR represent the 
synthesized beam width and signal-to-noise ratio, respectively). 

In summary, we conclude that the astrometric accuracy in 
our observations is mainly limited by the structure of the maser features, 
and that the atmospheric zenith delay residuals also contribute to some 
extent to the error sources. As suggested by \citet{honma2008b}, 
tropospheric zenith delay residual also plays as a random error because 
they are correlated only over a few days. Thus, one can reduce the post-fit 
residuals by employing longer observing epochs as well as using larger 
number of masers. Thus, the longer monitoring observatios will be helpful 
to achieve higher precision and accuracy of the parallax measurements, 
which is still underway.  

\subsection{Distance to IRAS~22198+6336 in L1204G}

We derived the distance to L1204G to be 764$\pm$27~pc from the Sun. 
This value is smaller than the earlier photometric result, 910~pc \citep{crampton1974}, 
and the kinematic distance, 1.3~kpc \citep{wouterloot1989, molinari1996}. 
While they did not report the errors in the derived distances, 
photometric and kinematic distances would generally contain larger uncertainties 
by a factor of a few tens of percent than the parallactic distances 
(e.g. Hirota et al. 2008, Sato et al. 2008, and references therein).  
The closer distance, 615$\pm$35~pc, is reported by \citet{dezeeuw1999} based on the 
Hipparcos measurements of the annual parallaxes of the Cepheus OB2 cluster members, 
in which one of the members HD211880 is the ionization source of S140 
that is possibly associated with L1204. 
This value is offset from our measured distance of IRAS~22198+6336 by 149$\pm$44~pc 
and is slightly larger than the apparent diameter of the Cepheus~OB2 cluster of 
about 130~pc ($\sim12$~degrees in the plane of the sky; de Zeeuw et al. 1999). 
If L1204G is physically associated with the Cepheus~OB2 cluster, 
our result suggests that L1204G would be located at the farthest side of 
the Cepheus~OB2 cluster. 
Our parallactic distance to IRAS~22198+6336 is found to be very close to 
the well studied H$_{2}$O maser sources in the Cepheus-Cassiopeia region 
such as IC1396N \citep{patel2000} and Cepheus~A \citep{torrelles2001}, 
although they are not based on the parallax measurements. 
Together with parallactic distances of the maser sources in this region including 
NGC281 reported by \citet{sato2008} and further measurements such as for 
IC1396N and Cepheus~A, 
we will be able to reveal the 3-dimensional structure of this 
molecular cloud complex and refine the statistical 
studies of star-formation in this region \citep{yonekura1997}. 

\subsection{Physical Properties of IRAS~22198+6336}

Although the adopted FIR luminosity of IRAS~22198+6336 is different from 
literature to literature, it has been included in the sample of massive YSOs 
judging from the relatively large far-infrared or bolometric luminosity 
\citep{palla1991, molinari1996, sanchez2008}. 
For instance, \citet{tafalla1993} adopted $L_{IRAS}$ to be 250$L_{\odot}$ 
while \citet{tofani1995} adopted $L_{FIR}$ to be 580$L_{\odot}$ 
by assuming the same distance $\sim$900~pc. In contrast, 
\citet{molinari1996} assumed the kinematic distance of 1.3~kpc and 
derived the larger FIR luminosity of 1240$L_{\odot}$. 
Very recently, \citet{sanchez2008} reported results of a survey of 
millimeter and centimeter continuum emission from high-mass to 
intermediate-mass YSOs including IRAS~22198+6336, of which they adopted 
the luminosity of 1300$L_{\odot}$. 
Based on the radio continuum observations with the VLA of IRAS~22198+6336 
along with the observed flux densities reported by previous literatures, 
\citet{sanchez2008} presented basic physical properties of the radio counterpart 
of the IRAS source, VLA~2, such as mass and luminosity. 

However, all of these results employed the larger distances 
than our present result. Here we refine the physical properties of the 
possible powering source of the H$_{2}$O masers, IRAS~22198+6336 VLA~2, 
based on our newly derived distance and previously reported flux values 
by multiplying a scaling factor proportional to an appropriate 
power of the distance $D$. 
For instance, a factor of (764~pc/1300~pc)$^{2}$=0.35 is adopted for 
correcting the luminosity reported by \citet{sanchez2008}. 

First of all, we stress that the bolometric luminosity 
is reduced to be 450$L_{\odot}$, which is only 35\% of previously adopted value, 
1300$L_{\odot}$. 
Such a low luminosity would argue against a classification of IRAS~22198+6336 
as a massive YSO.
The derived luminosity set limits on the stellar mass of IRAS~22198+6336. 
If it is a zero-age main sequence (ZAMS) star, 
the spectral type of late-B (between B6 and B7) with the stellar mass of 
about 5$M_{\odot}$ \citep{stahler2004} well accounts for the derived luminosity. 
On the other hand, if it is in a pure accretion phase, the stellar mass 
of 7$M_{\odot}$ is derived from the accretion luminosity with 
the typical mass accretion rate of 10$^{-5}M_{\odot}$~yr$^{-1}$ and 
the stellar radius of 5$R_{*}$ \citep{stahler2004}. 
The newly obtained total (gas+dust) mass of IRAS~22198+6336, (7-10)$M_{\odot}$, 
which is only 35\% of that derived from the SED fitting 
\citep{sanchez2008}, is in good agreement with the above values. 
Thus, our result implies that IRAS~22198+6336 is not a massive 
but an intermediate-mass YSO with the mass of no more than 7$M_{\odot}$. 
The lack of the IRAS~12~$\mu$m emission as well as no near-infrared counter 
part in the 2MASS images support the hypothesis 
that IRAS~22198+6336 is deeply embedded intermediate-mass YSO, which is 
analogous to low-mass Class~0 source \citep{sanchez2008}. 

According to \citet{sanchez2008}, IRAS~22198+6336 is associated with 
centimeter radio continuum emission, which could arise either in 
an ultra-compact H{\sc{ii}} region or a shock-ionized region in the outflow. 
They concluded that the contribution of shock-ionized gas around IRAS~22198+6336 
should be only a small fraction of the observed centimeter emission because of 
an extremely high ratio of the centimeter continuum luminosity $S_{\nu} D^{2}$ to 
the outflow momentum rate, $\dot P$, with an efficiency factor of 90\%. 
However, if we assume the distance of 764~pc, the ratio is reduced by a factor 
of 0.59 because the luminosity is proportional to $D^{2}$ while the outflow momentum rate, 
which is usually derived from the linear size, radial velocity, and inclination angle of the 
outflow, is proportional to $D$. Even if this is the case, the efficiency of 50\% is 
still higher than the typical low-mass YSOs, $\sim$10\% \citep{anglada1995}, 
suggesting that it would be significantly 
higher in the very early-stage of intermediate-mass YSOs. 

The mass and luminosity of IRAS~22198+6336 refined by our newly derived distance 
is consistent with that of the intermediate-mass deeply embedded YSO with 
the spectral type of late-B. 
Nondetection of the 6.7~GHz CH$_{3}$OH masers, which is thought to be a signpost of 
massive star-formation site, toward IRAS~22198+6336 
\citep{szymczak2000} also supports our conclusion that IRAS~22198+6336 is 
not a massive YSO. Although the OH masers, which is another signature of massive 
star-formation associated with an ultra-compact H{\sc{ii}} region, are detected 
toward IRAS~22198+6336 \citep{edris2007}, 
relatively lower resolution of the OH maser observations, $\sim$3\arcmin, 
might give argument against their association with the same powering source of 
the H$_{2}$O masers. 
Further observational studies with the higher-resolution and higher-sensitivity 
centimeter, millimeter, and submillimeter interferometers 
would shed light on the properties of the central star in L1204G more clearly. 
Such detailed observational studies on IRAS~22198+6336 would be quite 
important because it is one of the rare sources in the very early stage of 
intermediate-mass YSO associated with the H$_{2}$O masers (e.g. Patel et al. 2000) 
and hence, good target sources for the complete 
understanding of the very early stage of intermediate-mass star-formation processes. 

\subsection{Kinematics of the Jet Traced by the H$_{2}$O masers in IRAS~22198+6336}
\label{sec-model}

Together with the annual parallax, we successfully measured the proper motions 
of the maser spots around IRAS~22198+6336. These results enable us to 
discuss in detail about the kinematics of the circumstellar materials. 
Prior to discussions, 
we should note that the proper motions obtained with VERA do not represent 
the "intrinsic" proper motions of the masers because they are measured with 
respect to the Sun and hence, include the contribution of the Solar motion 
and the Galactic rotation. 
Assuming the Solar motion relative to the LSR based on the Hipparcos satellite 
data, ($U_{0}, V_{0}, W_{0}$)=(10.00, 5.25, 7.17)~km~s$^{-1}$ \citep{dehnen1998}, 
we estimated the contribution of the Solar motion to the observed absolute 
proper motion to be 3.612~mas~yr$^{-1}$ and $0.064$~mas~yr$^{-1}$ in right ascension 
and declination, respectively. 
In addition, contributions of the Galactic rotation is calculated to be 
($-4.543$~mas~yr$^{-1}$, $-2.947$~mas~yr$^{-1}$) on the assumption of the flat rotation 
with $R_{0}$ of 8.0~kpc \citep{reid1993} and $\Theta_{0}$ of 236~km~s$^{-1}$ 
\citep{reid2004} for L1204G at the distance of 764~pc. 
Subtracting these values from the observed proper motions listed in 
Table \ref{tab-results} and shown in Figure \ref{fig-allspot}(b), we obtained 
the intrinsic proper motions of the H$_{2}$O masers corrected for the 
Solar motion and Galactic rotation. 

The origin of these proper motions is attributed to either 
due to the systemic motion of the host cloud L1204G itself, 
the internal motions of the masers with respect to the powering source 
IRAS~22198+6336, such as jets and rotating disk associated with it, or both. 
Regarding to the systemic motion, observed proper motions might trace the 
peculiar motion of L1204G with respect to the Galactic rotation. 
The discrepancy between the kinematic distance of L1204G, which is 
obtained by assuming that L1204G exactly follows the Galactic rotation curve, 
and our parallactic distance would be suggestive of the effect of 
the peculiar motion of L1204G rather than the Galactic rotation. 
Indeed, L1204G ($b\sim5.7$~degrees) is situated at 75~pc above 
the Galactic plane with a half-thickness of 87~pc \citep{dame1987}, 
which seems too large to apply the flat rotation model. 
However, the systemic motion of L1204G is quite uncertain because of a 
lack of the proper motion measurements of 
IRAS~22198+6336, as suggested in the previous result with VERA of NGC~1333 
\citep{hirota2008}. 
Therefore, here we only consider several possibilities for the origin 
of internal proper motions of the masers and the kinematics 
in the circumstellar material. 

As shown in Figure \ref{fig-allspot}(a), 
the proper motions of the features J-M provide a hint of systematic motion 
directed outward from the center of the circular structure with the 
approximate diameter of $\sim$300~mas or 200~AU. 
It is naturally interpreted that the most blue-shifted and 
red-shifted components trace the interacting region with the ambient gas 
and the collimated jets aligned in the east-west direction, and that 
the powering source is located at the center of the maser distribution 
(see Figure \ref{fig-allspot}(a)). This is roughly consistent with the position 
and elongation of the VLA source observed at the 7~mm band \citep{sanchez2008}, 
although their spatial resolution is much larger than our VLBI observations. 
The possible trend of the northward proper motion vectors in Figure 
\ref{fig-allspot}(b) could be due to the unknown systemic motion of the powering source. 
The close-up view of the main features shows an arc-like structure 
indicative of a bow shock, as seen in Figure \ref{fig-allspot}(b). 
Such bow shock structures at the tips of the collimated jets 
have been found in low-mass Class~0 protostars IRAS~05413-0104 
\citep{claussen1998} and S106~FIR \citep{furuya2000}. 

In this case, the origin of the other maser spots close to the systemic 
velocity, which seem to be distributed perpendicular to the possible jets, 
would be accounted for by the rotating disk. 
The radial velocity difference along the north-south direction, 
$\sim$8~km~s$^{-1}$ and their separation, $\sim200$~AU ($\sim$300~mas), implies 
the Keplerian rotation with the centrally enclosed mass of 1.6$M_{\odot}$/(cos~$i$)$^{2}$ 
where $i$ is the inclination angle between the rotation axis and the line-of-sight. 
The inclination angle of the disk $i$ can be estimated from that of the east-west 
jets traced by the most blue-shifted and red-shifted masers as follows. 
The mean proper motions of the most blue-shifted masers are derived to be 
$-1.53$~mas~yr$^{-1}$ 
and $3.15$~mas~yr$^{-1}$ in right ascension and declination, respectively. 
If we adopt the proper motions only in the right ascension direction as the 
outward velocity of the jets, the mean transverse velocity of 5.5~km~s$^{-1}$ 
toward west and the mean radial velocity of -6.6~km~s$^{-1}$ with respect to 
the systemic velocity for the maser spots J-M give the inclination angle $i$ 
of 50~degrees. 
Similarly, if we adopt the proper motions of both the right ascension and 
declination direction, the mean transverse velocity of 12.6~km~s$^{-1}$ 
yields the inclination angle $i$ of 66~degrees. 
The enclosed mass is estimated to be 3.9$M_{\odot}$ and 9.7$M_{\odot}$ 
for former and latter case, respectively, and these values are almost 
consistent with the result of the SED fitting $\sim 7M_{\odot}$. 

On the other hand, the radio continuum emission in the 1.3~cm and 7~mm bands 
showed elongated structure along the northwest-southeast direction \citep{sanchez2008}. 
According to their SED analysis, the 1.3~cm emission is dominated by the free-free 
emission from the shock-ionized jets while the 44\% of the 7~mm flux comes 
from the thermal dust emission in the circumstellar disk, and hence, 
the 7~mm emission would be due to a superposition of the ionized jet and the dust disk 
\citep{sanchez2008}. 
The direction of the radio continuum emissions are consistent with that 
of the large-scale velocity gradient \citep{tafalla1993, jenness1995}, and hence, 
\citet{sanchez2008} attributed them to the free-free emission from the ionized radio jets. 
However, the large-scale velocity gradient showed opposite 
trend to our H$_{2}$O maser distribution as mentioned in section \ref{sec-distribution}. 
The maser distributions might trace an interface between the ambient material and the 
outflow lobe which is parallel to the radio continuum source (e.g. Moscadelli et a. 2006), 
whereas the asymmetric distribution of the masers with respect to the outflow axis 
cannot be accounted for by the simple model only.

It is worth considering another possibility of the rotating disk around the 
YSO with a rotation axis along the north-south direction. 
If we assume the radial velocity gradient along the east-west 
direction, $\sim$16~km~s$^{-1}$ per $\sim200$~AU ($\sim$300~mas), as 
a cause of rotation, the enclosed mass is estimated to be at least 
6.4$M_{\odot}$, which is comparable to the total mass or stellar mass of 
IRAS~22198+6336 as discussed in the previous section. 
In this case, the northern and southern components C-I would 
trace the jets from the central YSO. Their radial velocity close 
to the systemic velocity would indicate that the jets are aligned 
almost within the plane of the sky and hence, the disk is almost 
edge-on view. The directions of the radio continuum emissions in the 1.3~cm 
and 7~mm bands \citep{sanchez2008} are in good agreement with 
that of the H$_{2}$O maser features C-I. 
Nevertheless, this model cannot explain the observed 
velocity drift of 0.2~km~s$^{-1}$~yr$^{-1}$ 
for the $-20$~km~s$^{-1}$ feature \citep{valdettaro2002, brand2003} 
because the velocity drift at the edge of the disk 
(at the radius of 100~AU) 
is estimated to be only 0.01~km~s$^{-1}$~yr$^{-1}$ in the case 
of the YSO with the mass of 5$M_{\odot}$. 
Thus, the rotating disk with the axis of north-south direction 
is inadequate to explain the kinematics 
in the circumstellar materials around IRAS~22198+6336 probed by the 
blue-shifted maser features J-M. 

Alternatively, distributions and proper motions of the features J-M could also 
be interpreted as a part of the smaller-scale expanding shell as 
shown in Figure \ref{fig-allspot}(b), with the radius of 29.1$\pm$0.8~mas 
(22.2$\pm$0.6~AU) and the expanding velocity of 3.8$\pm$0.4~km~s$^{-1}$ 
in the plane of the sky. In this case, 
the dynamical time of the expanding shell is calculated to be only 8~years. 
Basic characteristics of this expanding shell is quite similar to 
the case of Cepheus~A \citep{torrelles2001} and W75N \citep{torrelles2003}. 
Meanwhile, the nature of the possible powering source of the shell 
is not consistent with the position and elongation of the radio 
continuum emission \citep{sanchez2008}.  In addition, 
it is unlikely that such a shell is traced only by the most blue-shifted 
components. Thus, we rule out the possibility of the small-scale 
expanding shell as a kinematical model of the blue-shifted maser features J-M. 

\citet{brand2003} proposed three possible mechanisms to account for 
the velocity drift of the maser emission; a rotating circumstelar disk, 
acceleration/deceleration of the jets, or a precession of the jets. 
According to our discussion, most plausible explanation for the 
proper motions, radial velocity distribution, and velocity drifts 
of the maser features is the collimated bipolar jets along the east-west 
direction. Thus, the origin of the velocity shift would be 
the deceleration (because the radial velocity of 
the blue-shifted features became closer to the systemic velocity) 
or precession of the jets. 

To distinguish between possible kinematic models, 
proper motion measurements of red-shifted components located at 
the eastern end of the maser feature distributions would be crucial. 
In addition, further high resolution observations of centimeter to 
submillimeter continuum emission as well as near- and mid-infrared 
observations of the central YSO and collimated jets, if they are associated, 
would help to reveal the overall circumstellar structure of this source 
with accuracies of much better than 0.1\arcsec. 

\section{Conclusion}

In this paper, we presented the results of multi-epoch VLBI astrometry 
with VERA of the 22~GHz H$_{2}$O masers associated with IRAS~22198+6336 in L1204G. 
Using the absolute positions of total 26~maser spots that were detected 
in at least three observing sessions, we derived the annual 
parallax of IRAS~22198+6336 to be 1.309$\pm$0.047~mas, corresponding 
to the distance of 764$\pm$27~pc from the Sun. 
This is the most accurate distance to L1204G with the uncertainty of only 4\%. 

Because the newly derived distance is closer than those well adopted 
in the previous literatures (910~pc or 1.3~kpc), the bolometric luminosity 
of IRAS~22198+6336 should be significantly reduced. 
According to the SED of IRAS~22198+6336, 
the bolometric luminosity and total mass of IRAS~22198+6336 were 
refined to be 450$L_{\odot}$ and 7$M_{\odot}$, respectively. 
These values are not consistent with a massive YSO but are in agreement with 
an intermediate-mass YSO. We confirmed 
that IRAS~22198+6336 would be a deeply embedded intermediate-mass YSO analogous to 
a low-mass Class~0 source \citep{sanchez2008}. 

Together with the annual parallax, we measured the absolute proper motions of 
the H$_{2}$O masers. 
We discussed several kinematical models of the circumstellar gas such as 
a small scale expanding shell, collimated jets, or rotating disk. 
We proposed that the distributions, proper motions, and the previously 
reported velocity drifts 
\citep{valdettaro2002, brand2003} 
of the maser features associated with IRAS~22198+6336 
are consistent with the collimated bipolar jets and possibly rotating disk 
perpendicular to them. 

IRAS~22198+6336 is one of the rare sources in the very early stage of 
intermediate-mass YSOs. 
Therefore, further high-resolution and high-sensitivity 
observations of the continuum emission from centimeter to submillimeter wavelength, 
in particular with the future interferometers such as EVLA and ALMA, 
and the proper motion measurements of the H$_{2}$O masers and continuum sources 
will be essential  
to reveal the kinematics and more detailed properties of IRAS~22198+6336, 
which will contribute to the understanding of very early stage of 
intermediate-mass star-formation processes. 

\vspace{12pt}
We thank the anonymous referee for valuable comments and suggestions. 
We also thank Guillem Anglada and Carlos Carrasco-Gonzalez for providing us 
with information on the VLA observations of IRAS~22198+6336. 
We are grateful to the staff of all the VERA stations for their assistance in observations. 
TH is financially supported by Grant-in-Aids for Scientific Research 
from The Ministry of Education, Culture, Sports, Science and Technology 
(13640242, 16540224, and 20740112).


\begin{thebibliography}{}
\bibitem[Anglada(1995)]{anglada1995}
  Anglada, G. 1995, Rev. Mexicana Astron. Astrof. Ser. Conf., 1, 67
\bibitem[Brand et al.(2003)]{brand2003}
  Brand, J., Cesaroni, R., Comoretto, G., Felli, M., Palagi, F., 
  Palla, F., \& Valdettaro, R. 2003, A\&A, 407, 573
\bibitem[Chikada et al.(1991)]{chikada1991} Chikada, Y., et al. 
  1991, in Frontiers of VLBI, ed. H. Hirabayashi, M. Inoue, \& H. Kobayashi 
  (Tokyo: Universal Academy Press), 79 
\bibitem[Claussen et al.(1998)]{claussen1998} Claussen, M. J., Marvel, K. B., 
  Wootten, A., \& Wilking, B. A. 1998, ApJ, 507, L79
\bibitem[Crampton \& Fisher(1974)]{crampton1974} Crampton, D. \& Fisher, W. A. 
  1974, Pub. Dom. Astrophys. Obs., 14, 283
\bibitem[Dame et al.(1987)]{dame1987} Dame, T. M., et al. 1987, ApJ, 322, 706
\bibitem[Dehnen \& Binney(1998)]{dehnen1998} Dehnen, W. \& Binney, J. J. \ 1998, 
   MNRAS, 298, 387
\bibitem[de Zeeuw et al.(1999)]{dezeeuw1999} 
   de Zeeuw, P. T., Hoogerwerf, R., de Bruijne, J. H. J., Brown, A. G. A., 
   Blaauw, A. \ 1999, AJ, 117, 354
\bibitem[Edris et al.(2007)]{edris2007}
  Edris, K. A., Fuller, G. A., \& Cohen, R. J. 2007, A\&A, 465, 865
\bibitem[Fukui(1989)]{fukui1989} Fukui, Y. 1989, in Low-mass Star 
  Formation and Pre-Main-Sequence Objects, ed. B. Reipurth (Garching: ESO), 95
\bibitem[Furuya et al.(2000)]{furuya2000}
   Furuya, R. S., Kitamura, Y., Wootten, H. A., Claussen, M. J., Saito, M., Marvel, K. B., 
   \& Kawabe, R. 2000, ApJ, 542, L135
\bibitem[Hachisuka et al. (2006)]{hachisuka2006} 
  Hachisuka, K. et al. \ 2006, ApJ, 645, 337
\bibitem[Hirota et al.(2007)]{hirota2007} Hirota, T. et al. \ 2007, PASJ, 59, 897
\bibitem[Hirota et al.(2008)]{hirota2008} Hirota, T. et al. \ 2008, PASJ, 60, 37
\bibitem[Honma et al.(2003)]{honma2003} Honma, M. et al. \ 2003, PASJ, 55, L57
\bibitem[Honma et al.(2007)]{honma2007} Honma, M. et al. \ 2007, PASJ, 59, 889
\bibitem[Honma et al.(2008a)]{honma2008a} Honma, M. et al. \ 2008a, submitted to PASJ
\bibitem[Honma et al.(2008b)]{honma2008b} Honma, M., Tamura, Y., \& Reid, M. J. \ 2008b, PASJ, 60, in press
\bibitem[Iguchi et al.(2005)]{iguchi2005} Iguchi, S., Kurayama, T., 
  Kawaguchi, N., Kawakami, K. \ 2005, PASJ, 57, 259
\bibitem[Imai et al.(2007)]{imai2007} Imai, H. et al. \ 2007, PASJ, 59, 1107
\bibitem[Jenness et al.(1995)]{jenness1995}
  Jenness, T., Scott, P. F., \& Padman, R. 1995, MNRAS, 276, 1024
\bibitem[Kawaguchi et al.(2000)]{kawaguchi2000} Kawaguchi, N., Sasao, T., Manabe, S. \ 2000, 
  in Proc. SPIE, 4015, Radio Telescopes, ed.\ H. R. Butcher (Washington: SPIE), 544
\bibitem[Loinard et al.(2005)]{loinard2005}
  Loinard, L., Mioduszewski, A. J., Rodr\'\i guez, L. F., Gonz\'alez, R. A., 
  Rodr\'\i guez, M. I., Torres, R. M. \ 2005, ApJ, 619, L179
\bibitem[Loinard et al.(2007)]{loinard2007} 
  Loinard, L. Torres, R. M., Mioduszewski, A. J., Rodr\'\i guez, L. F., 
  Gonz\'alez-L\'opezlira, R. A., 
  Lachaume, R., V\'azquez, V., \& Gonz\'alez, E. 2007, ApJ, 671, 546
\bibitem[Loinard et al.(2008)]{loinard2008}
  Loinard, L., Torres, R. M., ; Mioduszewski, A. J., \&   Rodr\'\i guez, L. F. 
  2008, ApJ, 675, L29
\bibitem[Lombardi et al.(2008)]{lombardi2008} 
  Lombardi, M., Lada, C. J., \& Alves, J. 2008, A\&A, 480, 785
\bibitem[Menten et al.(2007)]{menten2007}  Menten, K. M., Reid, M. J., Forbrich, J., \& 
   Brunthaler, A. 2007, A\&A, 474, 515
\bibitem[Migenes et al.(1999)]{migenes1999}
   Migenes, V. et al. 1999, ApJS, 123, 487
\bibitem[Molinari et al.(1996)]{molinari1996} Molinari, S., Brand, J., 
   Cesaroni, R., \& Palla, F. 1996, A\&A, 308, 573
\bibitem[Moscadelli et al.(2006)]{moscadelli2006}
  Moscadelli, L., Testi, L., Furuya, R. S., Goddi, C., Claussen, M., Kitamura, Y., 
  \& Wootten, A. 2006, A\&A, 446, 985
\bibitem[Palla et al.(1991)]{palla1991} Palla, F., Brand, J., 
  Cesaroni, R., Comoretto, G., \& Felli, M. 1991, A\&A, 246, 249
\bibitem[Patel et al.(2000)]{patel2000} 
  Patel, N. A., Greenhill, L. J., Herrnstein, J., Zhang, Q., Moran, J. M., 
  Ho, P. T. P., \& Goldsmith, P. F. 2000, ApJ, 538, 268
\bibitem[Perryman et al.(1997)]{perryman1997}
  Perryman, M. A. C., et al. \ 1997, A\&A, 323, L49
\bibitem[Petrov et al.(2005)]{petrov2005} 
  Petrov, L., Kovalev, Y. Y., Fomalont, E., \& Gordon, D. 2005, AJ, 129, 1163
\bibitem[Reid(1993)]{reid1993} Reid, M. J. \ 1993, ARA\&A, 31, 345
\bibitem[Reid \& Brunthaler(2004)]{reid2004} Reid, M. J. \& Brunthaler, A. \ 2004, 
   ApJ, 616, 872
\bibitem[S\'anchez-Monge et al.(2008)]{sanchez2008} S\'anchez-Monge, \'A., Palau, A., Estalella, R., 
  Beltr\'an, M., \& Girart, J. M. 2008, A\&A, 485, 497
\bibitem[Sandstrom et al.(2007)]{sandstrom2007} Sandstrom, K. M., Peek, J. E. G., 
  Bower, G. C., Bolatto, A. D., \& Plambeck, R. L. 2007, ApJ, 667, 1161
\bibitem[Sato et al.(2007)]{sato2007} Sato, M. et al. \ 2007, PASJ, 59, 743
\bibitem[Sato et al.(2008)]{sato2008} Sato, M. et al. \ 2008, PASJ, 60, in press
\bibitem[Stahler \& Palla(2004)]{stahler2004}
   Stahler, S. W. \& Palla, F. 2004,  The Formation of Stars (Weinheim: Wiley-VCH)
\bibitem[Szymczak et al.(2000)]{szymczak2000}
  Szymczak, M., Hrynek, G., \& Kus, A. J. 2000, A\&AS, 143, 269
\bibitem[Tafalla et al.(1993)]{tafalla1993} Tafalla, M., Bachiller, R., \& 
  Mart\'\i n-Pintado, J. 1993, ApJ, 403, 175
\bibitem[Tofani et al.(1995)]{tofani1995} Tofani,G., Felli,M., Taylor, G. B., \& Hunter, T. R. 
  1995, A\&AS, 112, 299
\bibitem[Torrelles et al.(2003)]{torrelles2003} Torrelles, J. M. et al. 2003, ApJ, 598, L115
\bibitem[Torrelles et al.(2001)]{torrelles2001} Torrelles, J. M. et al. 2001, Nature, 411, 277
\bibitem[Torres et al.(2007)]{torres2007}
  Torres, R. M., Loinard, L., Mioduszewski, A. J., \& 
  Rodr\'\i guez, L. F. 2007, ApJ, 671, 1813
\bibitem[Ulich \& Haas(1976)]{ulich1976}
   Ulich, B. L., Haas, R. W. \ 1976, ApJS, 30, 247
\bibitem[Valdettaro et al.(2002)]{valdettaro2002}
  Valdettaro, R., Palla, F., Brand, J., Cesaroni, R., Comoretto, G., 
  Felli, M., \& Palagi, F. 2002, A\&A, 383, 244
\bibitem[Wilking et al.(1989)]{wilking1989}
  Wilking, B. A., Blackwell, J. H., Mundy, L. G., \& Howe, J. E. 
  1989, ApJ, 345, 257
\bibitem[Wouterloot \& Brand(1989)]{wouterloot1989} 
  Wouterloot, J. G. A. \& Brand, J. 1989, A\&AS, 80, 149
\bibitem[Yonekura et al.(1997)]{yonekura1997} 
   Yonekura, Y., Dobashi, K., Mizuno, A., Ogawa, H., \& Fukui, Y. 1997, ApJS, 110, 21
\end{thebibliography}
\end{document}